# Identifying Research Fields within Business and Management: A Journal Cross-Citation Analysis

**John Mingers (**Corresponding author)
Kent Business School, University of Kent, Canterbury CT7 2PE, UK
phone: +44 (0)1227 824008
fax: +44 (0)1227 76118
e-mail: j.mingers@kent.ac.uk

**Loet Leydesdorff**
Amsterdam School of Communication Research (ASCoR),
University of Amsterdam, 1012 CX Amsterdam, The Netherlands
e-mail: loet@leydesdorff.net .

## Abstract

A discipline such as business and management (B&M) is very broad and has many fields within it, ranging from fairly scientific ones such as management science or economics to softer ones such as information systems. There are at least two reasons why it is important to identify these sub-fields accurately. Firstly, for the purpose of normalizing citation data as it is well known that citation rates vary significantly between different disciplines. Secondly, because journal rankings and lists tend to split their classifications into different subjects – for example, the Association of Business Schools (ABS) list, which is a standard in the UK, has 22 different fields. Unfortunately, at the moment these are created in an *ad hoc* manner with no underlying rigour. The purpose of this paper is to identify possible sub-fields in B&M rigorously based on actual citation patterns. We have examined 450 journals in B&M which are included in the ISI Web of Science (WoS) and analysed the cross-citation rates between them enabling us to generate sets of coherent and consistent sub-fields that minimise the extent to which journals appear in several categories. Implications and limitations of the analysis are discussed.

# 1 Introduction

Business and Management (B&M) constitutes a wide and disparate research area. Its boundaries with other disciplines are fuzzy, both because it draws on a range of foundational disciplines and because it has many application areas. It is also complex within itself, having different sub-disciplines, application areas and technologies. In this paper we will consider the latter problem and attempt to identify a group of clearly demarcated sub-fields within B&M as a whole. Why is this a useful thing to do? Two particular reasons concern research evaluation using citations, and the increasing importance of journal ranking lists such as the one created by the Association of Business Schools (ABS) (Association of Business Schools, 2010) in the UK.

Considering firstly citations, it is increasingly the case that research evaluation is being carried out through bibliometric analysis based on citations, either instead of or combined with peer review. It is clear through many empirical studies (Leydesdorff, 2008; Mingers and Burrell, 2006; Moed et al., 1985; Rinia et al., 1998) that citation behaviour, in terms of the average number of citations per paper, varies dramatically between different disciplines (as well as depending on other factors such as age of paper, type of paper and journal). Generally, the sciences cite much more highly than the social sciences, which in turn cite more highly than the humanities but within each of the areas there are also wide dispersions. This means that, in comparative analyses, whether at the level of the individual researcher, the research unit, or the journal, citation data must be normalised to the field of study. This clearly requires that there exists an agreed set of fields or sub fields, each with its own collection of journals, in order to do the normalisation. However, no such sets exist at the moment. One can question whether all journals can unambiguously be attributed to disciplines or specialties (Boyack and Klavans, 2011).

Most citation analyses use one of the major citation databases such as Thomson-Reuters Web of Science (WoS) or Elsevier’s Scopus. One of the main centres for this type of research evaluation is the Centre for Science and Technology Studies (CWTS) at Leiden University (Moed, 2010; van Raan, 2003; van Raan et al., 2011). They have developed their own methodology – the Leiden Ranking Methodology – based on citations taken from the WoS. For the purposes of normalisation, they rely on the definitions of fields within WoS. Whilst it may be reasonable for other disciplines, it is certainly not for B&M (Mingers and Lipitakis, 2013). Table 1 show the three main fields relevant to B&M – Management, Business, Business Finance, together with several others that are also relevant. The first problem is that these fields are not defined clearly nor are they based on any underlying analysis (Pudovkin and Garfield, 2002; Rafols and Leydesdorff, 2009). What exactly is the difference between the three? On looking at the journals within them, they cover what are seen within B&M as very different sub-disciplines. In comparison, the ABS journal list, which we will discuss below, has 22 different categories within it. Of the other related ones, "OR and management science" is actually listed in the Science database rather than the Social Science one; information systems is combined with library science; and the other two are somewhat eclectic.

As can be seen from the table, there is also a considerable degree of overlap with the same journal appearing on two or even three fields. This would not perhaps matter so much if the citation levels of the different fields were all similar, but in fact one of the characteristics of B&M is that it has a wide range of very diverse disciplines, from scientific ones such as operational research and economics, through social science ones like organisation studies, to soft, philosophical discourses. If a paper appears in more than one field (and of course some genuinely may do) and the fields have different normalisation rates, it is difficult to find a reasonable value.

Moving to journal ranking lists, they are assuming increasing importance in the assessment of research quality. It is extremely time consuming, and far from objective, to judge the quality of every published paper by peer review. It is therefore very common, instead, to use the supposed quality of the journal as a proxy for the quality of its papers which displaces the problem to assessing the journal quality, hence the use of journal ranking lists. The use of journal rankings in this way is of course contentious. (2007, 2008), who was a member of the 2008 RAE Panel, states that "One major conclusion appears to be that journal rankings are not a good indicator of the quality of any paper published in that journal, nor necessarily the combined quality of all the papers" (Paul, 2008, p. 324). Macdonald and Kam (2007) argue that the of academic publishing in management is one of gamesmanship and game playing. Adler and Harzing (2009) provide another strong critique of the dysfunctional effects of academic ranking systems and journal rankings in particular. The main complaint is that they lead to a narrowing of the discipline, concentrating research into the narrow confines of established journals and discouraging innovation and interdisciplinary work (Rafols et al., 2012).

Within the UK, the regime of Research Assessment Exercises (RAEs), the current one (2013) being called the Research Excellence Framework (REF) (RAE, 2004, 2005, 2006), has placed huge emphasis on journal quality as business school Deans are faced with difficult decisions about which people and which papers to enter in their submissions to the REF. This has led to one particular journal list – the ABS one – becoming the *de facto* standard. It is clearly and explicitly used by all business school but it is also implicitly used by the REF Panel itself. Although they say publically that journal ranking lists will not be formally used, the sheer volume of papers to be assessed by a relatively small Panel makes it a necessity. In 2008, the Panel claimed that "most outputs were read in considerable detail" (RAE, 2009, p. 5) but this must have been an impossible task given that there were 12,600 papers to be read by 18 academics in only a few weeks (Mingers et al., 2012).

The ABS list itself has been extensively critiqued (Hoepner and Unerman, 2009; Hussain, 2011; Mingers and Willmott, 2013; Willmott, 2011) and defended (Morris et al., 2009; Morris et al., 2011). The list (currently version 4) covers 823 journals split into 22 categories. This seems a lot, but in fact papers in more than 1600 different journals were submitted in the last RAE and a lack of coverage of particular journals is one of the criticisms. Others are: i) that the categories are somewhat arbitrary and not based on an underlying rationale. ii) That the quality levels assigned to different categories are highly variable. For example, 16 out of 38 (42%) psychology journals are awarded the top 4 grading while only 2 out of 53 (4%) information systems journals were. iii) That in some categories (but not others), e.g., operational research, there is a bias towards US journals which exclude certain types of research of importance in the UK (e.g., soft OR). iv) That the process of compiling the list is not transparent and that the compilers of the list do not engage with subject communities.

In this paper we are mainly concerned with the first issue – that of the subject categories. They are shown in Table 2. As can be seen, there are quite a large number; they differ significantly in size (from 10 to 134 journals); and there is little justification for them. As one of the founders of the ABS list has said:

“The twenty two so-called subject fields in the *ABS Guide* are an eclectic mix of categories consisting of: academic disciplines (Business History; Economics; Organization Studies; Operations Research and Management Science; and Psychology); business functions (Accounting; Finance; Human Resource Management and Employment Studies; Information Management; Marketing; Operations and Technology Management; and Business Strategy); industries (Tourism and Hospitality Management); sectors (Entrepreneurship and Small Business; International Business and Area Studies; Public Sector Policy, Management and Administration; and Sector Studies, covering a wide range of specialisms that includes health and education); issues or interests (Ethics and Governance; Innovation and Technology Management; Management and Education); as well as more or less residual categories (General Management, which includes many of the leading business and management journals; and Social Sciences)” (Rowlinson, 2013).

Whilst it may be necessary that such a disparate field as business and management does require different kinds of sub-fields, it should be possible to generate them on the basis of actual publication and citation behavior rather than purely ad-hoc judgement.

This brings us to the subject of the paper. For the two reasons outlined, it would be valuable if a set of sub-fields could be identified in terms of journals within business and management. The method used here is to look at the actual citation and referencing behaviour of researchers in terms of the cross-citations between different journals. Given a matrix of the cross-citations between a large number of journals it should be possible to use statistical methods to discover patterns of cross-citation which essentially correspond to the sub-fields. In Section 2 we explain the data collection and statistical methods used. In Section 3 we present the results, and then in Section 4 we will discuss the implications and limitations of the study.

# 2. Data and Methodology

The data collected on citations came from the Journal Citation Reports at Thomson-Reuters' Web of Science which is the most reliable source of citations although it is limited in its coverage, especially in business and management (Mingers and Lipitakis, 2010). All the journals in the ABS list that are in WoS are classified with their ISI abbreviation and this was used to interrogate the WoS in order to obtain the number of citations from those journals, in 2011, to papers from those same journals over all years. This generated a matrix of citations in which the rows (observations) were the citing journals in 2011 and the columns (variables) were the cited journals across all the years. After cleaning, there were 453 variables and 449 cases. As is usual with cross-citation data (Leydesdorff, 2004) the matrix was very sparse with over 85% zeros. This dataset used only the most recent year's worth of data (2011) but there is little point in using more years unless one is doing longitudinal research to detect changes, which was not the purpose of this study.

Three different analysis techniques were used: the Blondel algorithm (2008) for constructing communities or groups from large data networks, traditional cluster analysis, and factor analysis (Zhao and Lin, 2010). The Blondel algorithm is a relatively recent heuristic that has been shown to be highly effective in analysing very large networks. It uses a measure of the modularity of a particular partition and works in two phases that are repeated iteratively. The first phase tests if modularity can be improved by swapping nodes between clusters; the second phase takes the clusters and treats them as the nodes of the network to which phase one is applied again. When applied to our dataset, nine clusters were generated but they were not very satisfactory. Five of them were quite large with between 30 and 130 journals in each, but the remaining four were very small with between one and seven journals in each. Moreover, the large groups included quite diverse communities, for example psychology, HR, marketing and management were all in the same grouping.

The next approach was traditional, agglomerative cluster analysis. In terms of method, two decisions have to be made: which measure of distance between nodes/clusters, and which agglomeration method should be used. With regard to distance measures, there are a range of possibilities based either on the Euclidian distance of the Pearson correlation coefficient. However, our data is quite unusual in that it has a large number of zeros and also a high degree of dispersion of values. Ahlgren, Jarneving and Rousseau (2003) found that Pearson's coefficient was inappropriate in these circumstances (in particular, simply adding in zero entries into the matrix changes the value), and that the Salton's (1987) cosine normalisation measure was more satisfactory (Egghe and Leydesdorff, 2009). For the agglomerative measures, all measures have their own particular biases in terms of the types of clusters that they form but Ward's method is considered very reliable. The other major question is how many groups to have where there is no theoretical reason for there being a specific number. It is possible to look at a scree diagram to see if there is a significant change of slope, but if there is not it is a matter of judgement based on knowledge of the domain and the coherence of the groups that have been formed. We can see from Tables 1 and 2 that the WoS have 8 relevant but overlapping groups, which the ABS list has 22. These could perhaps be seen as upper and lower limits, although certainly for citation normalisation purposes 22 is very high.

In the event we performed two cluster analyses based on cosine normalisation and Ward's method with 10 and 15 groups respectively. In both solutions there are several well-established and stable groupings – information systems/information technology, operations research/operations management, agricultural economics/development, psychology, economics and marketing. There are also some groups that get combined together, e.g., accounting and finance, and transport and regional. But, in both solutions there is one very large and very mixed cluster with 154 journals in the 15-group and 186 in the 10-group. This includes finance, health, technology, statistics, tourism, education, economics, HR and so on. Other clustering algorithms were tried but the results were broadly the same. These results were not considered satisfactory, and so the third analysis method – factor analysis, which has been recommended for this type of analysis (Leydesdorff, 2004, 2006), was deployed.

Factor analysis is a multivariate method that aims to uncover general factors that underlie a set of data with many variables (Hair et al., 1998). It is based on the correlations (or covariances) between variables. If all the variables were independent of each other, then each variable would be its own factor. But where there is a correlation structure we can explore the extent to which that is reflective of some underlying, or latent, factors. In our case, there is a pattern in the data in that the cited journals will tend to cluster as a result of the citing patterns of behaviour (of the same journals). We might expect that the journals will group into fields, and the factor analysis should be able to uncover what these fields are. There will be some journals that span several fields, and others that are very specialised to a particular field.

There are generally two stages in factor analysis – the extraction of the factors, and then the possible rotation of them. The most common extraction method is principal components analysis (PCA). This is an analytical method of data reduction that represents the variability (covariance) of a data set by extracting a set of orthogonal (independent) components in order of the amount of variability explained. The first PC is the linear combination of variables that captures the greatest amount of variability. It is similar but not identical to a regression line. The second factor is the line, orthogonal to the first, which captures the next greatest amount of variance. The process continues until there are as many components as variables and all the variance has been explained. In practice, one stops after a specified number of components have been extracted. This process means that each component is independent of the others so choosing to extract more components does not change the preceding ones. It also means that a decision has to be made about how many to extract. This can be based on theoretical considerations, or on the pattern of variance that is explained as more factors are extracted. There is potential a second process called rotation where the whole set of components can be rotated in multi-dimensional space in order to clarify the results – i.e., to make the components sharper. This rotation may be orthogonal (maintaining the independence of the components) or oblique.

In this analysis, the aim is to see if a relatively small set of underlying components, citing sub-fields, can explain the overall covariability of cited journals. PCA was used to extract the components, and two rotational methods were tried. The results, described in the next section, were very interesting.

## 3. Factor Analysis results

There were 453 cited journals that constituted the set of variables for this analysis. After the initial PCA extraction we need to consider if greater clarity can be obtained by rotating the factors. We considered only orthogonal rotations and there are two main types. The first, varimax, aims to simplify the columns of the factor loadings. That is to try and make the coefficients in each factor as near to 0 or |1| as possible. Alternatively, quartimax aims to make the coefficients for each row (in this case journal) as near to 0 or |1| as possible so that each variable is as clearly represented in only a small number of factors. In our case, the first approach tries to make each sub-field as clear as possible, with potentially a relatively small number of journals, but journals may appear in several sub-fields. The latter approach tries to link a particular journal to only one sub-field thus reducing the number of journals appearing in multiple sub-fields. Given that one of the purposes of the research was to avoid the problem of journals appearing in multiple fields, it was felt that quartimax was most suitable.
Table 3 shows the extraction details for the first 22 components. The first component had a variance (Eigenvalue) after rotation of 27.11 which by itself represented 5.99% of the total variance. The main question at this stage is how many components to retain. Statistical guidance suggests components with an Eigenvalue of greater than 1 which would give 21 components explaining 41.21% of the original variation. An examination of the scree plot does not show any significant points of discontinuity. However, we believe that it is better to consider this in terms of the actual classifications generated rather than just the statistical results.

The actual factor loadings table, with 453 journals and 22 factors is too large to present in the paper but is available on the publishers website. The method, however, works well generating groups that are generally clearly defined. The first two columns of Table 4 show brief descriptions of the groups together with the number of journals within them. Journals are allocated to the factor for which their loading is highest positively. They may also have significant loadings in other factors, indicating they are also well cited in other groups. They may also be negatively loaded on a factor indicating that there are less cross citations with journals in that factor then would be expected.

The 22-group solution covers 423 of the 453 journals in the data set (see Table 5). Those not included did not load significantly on these particular factors. These tend to be journals in specialised areas that would generate a factor of their own if more factors were extracted. For example, one group is seven education journals which, upon further analysis, were contained by principal component 54. The groups themselves do seem to have logical coherence and are a mix of disciplines, e.g., economics or OR, and application areas, e.g., energy and environment or transport. Comparing these groups with the 22 ABS ones, there are some clear differences, but we should remember that we are only dealing with a specific subset of ABS journals – those included in ISI Web of Science – and these are not distributed evenly across the ABS groups. For instance, over 70% of journals in economics, IS, OR, psychology and social science are included in ISI, while less than 30% of journals in accounting, ethics/governance, international business, management education and tourism are. Thus these latter categories are not well represented in our dataset.

The main ABS groups not included in our classification are: entrepreneurship, ethics, IB, innovation, management education and tourism as well as the catch-all categories of general management and sector studies which do not have coherence anyway. The groupings developed in our analysis that do not occur in ABS are mainly applied areas such as regional and environmental, energy, development and transport, although also appearing are more disciplinary areas such as statistics and informatics.

Overall, our classification is broadly similar to that of ABS but is more well-grounded in that it is based on actual citation patterns between journals rather than ad-hoc judgements. However it is subject to the limitation of poor coverage in ISI in certain areas particularly. Note that the position of a journal in the list in Table 5 is based purely on the loading of the journal into the group – i.e., the first ones are more central to the group than the later ones – but it does not imply anything about the *quality* of the journal.

We should perhaps discuss the split into two economics groups which is maintained in the results with less groups to be discussed below. At first sight it seems strange that there should be a split within economics, and looking at the journal titles does not display any particular clues. However, producing a graphical representation using network mapping software (Pajek/VOSViewer) (Figure 1) shows that actually there is a core of economics journals that are largely self-contained and that the second group actually cluster around the edge of the core ones. The implication is that the second group are journals that are more related to the rest of the B&M literature, as well as to the economics ones.

The aim of this research was not simply to replicate or improve on the ABS list. It was also concerned to produce a set of sub-fields that represented differential citation behaviour within the management discipline to improve normalisation processes and reduce the extent to which a journal was represented in several different groupings. With this in mind, and noting that WoS itself only has a small number of relevant sub-fields (no more than five or six), we went on to look at solutions with smaller numbers of factors and therefore groups. In particular, we will examine 10 and 15 group solutions. The statistical analysis actually moves from few groups to many groups as new ones are split off, but we will discuss it in the opposite direction. As the number of groups reduces, we find that three things can happen: i) groups move in their entirety into another group, examples being accounting into finance and informatics into information. Or ii) they spread across a small number of other groups, for example public administration into economics (periphery) and psychology. Or iii) they more or less disappear with journals being widely spread or not appearing significantly in any groups, for example statistics and economic history. It is noticeable from the column totals that the number of journals classified in the groups is reducing. This is because as the groups formed are larger, some journals no longer appear as significant within them. Or, in terms of the alternate direction, as more factors are produced, new groupings are generated and journals that were "lost on the crowd" now become significant.

Even in the 22-group analysis, 28 journals do not appear in any grouping. These are shown in Table 6. These can generally be seen to be peripheral to business and management as a whole, although some of them, *Ann Tourism Res* or *Hum Factors*, are slightly surprising. Table 6 also shows 15 journals that have significant loadings across at least seven different fields, indicating a high degree of cross-disciplinary material.

For the purposes of normalisation, what matters is whether different groups actually do differ significantly in terms of the number of citations they generate. To investigate this, we have calculated the mean citations per journal (for the year of our data – 2011) for each of the groups in all three solutions. These are also shown in Table 5. Beginning with the 10-group solution, we can see that the mean citations are very similar for the first six groups, being between 1599 and 1679. There is then economics (periphery) on 1169, energy on 938 followed by IS/IT and regional in the 600s. So at this level of resolution, one might wish to say that there are only two, or possibly three, groups that need to be differentiated. However, as the level of resolution increases, with more groups being separated out, the dispersion increases. To some extent this would be expected statistically – the fewer the groups (and thereby the larger), the more the means will tend towards the overall mean. But the results are really quite significant. For example, the management/strategy group grows from 1679 to 2279 to 2413 in the 22-group solution while IS/IT reduces from 609 to 343. The result in the 22-group case is three groups over 2000, seven groups over 1000, and twelve under 1000. Those at the top are five or six times greater than those at the bottom.

There are a few journals with extremely high citation counts, mainly in the first three groups, for example *Am Econ Rev* has 16000, *J Pers Soc Psychol* has 14000, *Man Sci*, *Acad Manage Rev*, J. *Finance* and *Econometrica* each have over 10000 which will affect the mean. But generally in these groups it is the large number of reasonably highly cited journals that generates the high mean. It might be suggested that the analysis method itself (factor analysis) might choose groups in terms of number of citations, but in fact the analysis was done on the correlation matrix rather than the covariance matrix and so was not affected by the absolute size of the citations. The main two factors generating the differences are: i) general differences in citation behaviour that are found between different disciplines, especially between sciences and the social sciences or humanities which might explain the high rates in economics, psychology, finance and OR; ii) size of population differences between general subjects and specialised or niche subjects. This might explain why, for example, the management/strategy category is high while public admin, development and transport are low – there are simply fewer academics writing and citing fewer papers in the specialist areas.

In terms of normalisation, it is certainly clear that there needs to be a differentiation between fields based on actual citation behaviour as opposed to the rather *ad hoc* groupings that currently exist in WoS. An ANOVA analysis suggests that there could be two groups – those above 1000 cites per journal and those below, but one could also suggest three groups – above 2000, above 1000 and below 1000. Further analysis of a larger set of journals would be needed to resolve this question more adequately.

# 4. Conclusions

This paper has shown that it is possible to identify sub-fields within the business and management discipline by analysing the cross-citations between journals. Using factor analysis, we have been able to construct several solutions, with different numbers of sub-fields, which are clear and consistent. There are two main reasons for doing this. The first is for the purpose of normalising citation metrics since citation rates vary significantly across disciplines. We have found that there are at least two significantly different groups of sub-fields with respect to citation rates whether we consider the 10-group or the 22-group solution. These are different from the fields that are defined in WoS, which are somewhat arbitrary, although they are often used for citation metrics.

The second reason is for journal ranking lists where the list as a whole needs to be split into a number of different subjects. The current ABS list has 22 and we have emulated that number although our groupings are different and have a more rigorous underlying logic.

The main limitation of this research is the set of journals that have been used as it does not fully represent the business and management literature. The sample is limited in two ways. First, because the citations were taken from WoS it only includes those journals in WoS and, as we have seen, there is a very uneven coverage across the different sub-fields. This will particularly affect the identification of sub-fields in those areas. The only way to overcome this is to use a different source of citations – in particular Google Scholar (GS) (Mingers and Lipitakis, 2010) – which covers all disciplines more evenly, although the citations themselves are less rigorously collected.

The second limitation is the ABS list itself which does not contain all journals within B&M. For example, in the 2008 RAE in the UK, papers from over 1600 journals were submitted to the B&M Panel, although some may well be in application disciplines rather than B&M itself.

**Acknowledgement**
We are grateful to Thomson-Reuters for permission to use the JCR data.

| | Business | Business Finance | Economics | Industrial Relations & Labor | Information Science & Library Science | International Relations | Management | Operational Research & Management Science |
|---|---|---|---|---|---|---|---|---|
| Business | 103 | - | 13 | - | - | 1 | 39 | - |
| Business Finance | - | 76 | 35 | - | - | 1 | 1 | - |
| Economics | 13 | 35 | 305 | 4 | - | 10 | 9 | 1 |
| Industrial Relations & Labor | - | - | 4 | 22 | - | - | 4 | - |
| Information Science & Library Science | - | - | - | - | 77 | - | 8 | - |
| International Relations | 1 | 1 | 10 | - | - | 78 | - | - |
| Management | 39 | 1 | 9 | 4 | 8 | - | 144 | 8 |
| Operational Research & Management Science | - | - | 1 | - | - | - | 8 | 75 |

**Table 1. Fields in WoS showing overlapping coverage (numbers of journals) from WoS 2011**

| Subject Code | Subject Covered | No. of journals |
|---|---|---|
| ACCOUNT | Accounting. This field includes auditing and taxation journals (See also Finance | 35 |
| BUS HIST | Business History. This field includes related specialist journals focusing on management, firms, industries and employees | 14 |
| . ECON | Economics. This is a very broad field with many sub-specialisms. The focus in the selection of journals has been on general economics journals and those that publish articles dealing with business, management and industrial economics and related fields. | 134 |
| ENT-SMBUS | Entrepreneurship and Small Business. | 17 |
| ETH-GOV | Ethics and Governance. | 16 |
| FINANCE | Finance. All general and specialist finance journals including insurance and actuarial journals. | 62 |
| GEN MAN | General Management. This is a broad field containing many of the"heartland" journals of business and management studies, which have a broad coverage and inter-disciplinary content. | 31 |
| HRM&EMP | Human Resource Management and Employment Studies. This field includes journals dealing with personnel, human resource management, employee and industrial relations as well as those that apply sociological perspectives to work and employment. | 35 |
| IB&AREA | International Business and Area Studies. This field brings together international business and interdisciplinary area studies. | 24 |
| INNOV | Innovation and technology change management. | 10 |
| INFO MAN | Information Management. Studies of information systems and information technology and information processes. | 53 |
| MGT&ED | Management and Education. This includes career, employee and management development as well as publications focusing on education, skills and training | 29 |
| MKT | Marketing. The field covers advertising and marketing and related. specialisms such as communications and public relations. | 54 |
| ORG STUD | Organization Studies. | 28 |
| OR&MANSCI | Operations Research and Management Science. This field includes the application of mathematical analysis, operations research, | 35 |
| OPS&TECH | Operations and Technology Management | 40 |
| PSYCH | Psychology. This is a small sub-set of the psychology journals that attract contributions from business and management academics. | 38 |
| PUB SEC | Public sector policy, management and administration | 33 |
| SECTOR | Sector Studies. This covers health, education, arts, not-for-profit, engineering and other fields of management practice. It extends beyond issues of services management to include specialisms in manufacturing and primary industries | 37 |
| SOC SCI | Social Sciences. These in the main are sociological, geographical economic historical, cultural and political journals that are attractive, publication outlets for business and management academics. | 60 |
| STRAT | . Business Strategy. | 12 |
| TOUR-HOSP | Tourism and Hospitality Management | 24 |

**Table 2 Subject Groups in ABS Journal List**

| Component | Initial Eigenvalues | | | Rotation Sums of Squared Loadings | | |
|---|---|---|---|---|---|---|
| | Total | % of Variance | Cumulative % | Total | % of Variance | Cumulative % |
| 1 | 32.481 | 7.170 | 7.170 | 27.114 | 5.985 | 5.985 |
| 2 | 20.193 | 4.458 | 11.628 | 14.557 | 3.213 | 9.199 |
| 3 | 14.663 | 3.237 | 14.865 | 12.695 | 2.802 | 12.001 |
| 4 | 11.859 | 2.618 | 17.482 | 12.296 | 2.714 | 14.715 |
| 5 | 10.273 | 2.268 | 19.750 | 10.097 | 2.229 | 16.945 |
| 6 | 9.309 | 2.055 | 21.805 | 9.283 | 2.049 | 18.994 |
| 7 | 8.419 | 1.858 | 23.664 | 9.090 | 2.007 | 21.000 |
| 8 | 7.980 | 1.762 | 25.425 | 8.874 | 1.959 | 22.959 |
| 9 | 7.488 | 1.653 | 27.078 | 8.558 | 1.889 | 24.848 |
| 10 | 7.184 | 1.586 | 28.664 | 7.549 | 1.666 | 26.515 |
| 11 | 6.780 | 1.497 | 30.161 | 7.374 | 1.628 | 28.143 |
| 12 | 6.473 | 1.429 | 31.589 | 7.290 | 1.609 | 29.752 |
| 13 | 6.036 | 1.332 | 32.922 | 6.957 | 1.536 | 31.288 |
| 14 | 5.398 | 1.192 | 34.113 | 6.593 | 1.455 | 32.743 |
| 15 | 5.109 | 1.128 | 35.241 | 6.097 | 1.346 | 34.089 |
| 16 | 4.920 | 1.086 | 36.327 | 5.684 | 1.255 | 35.344 |
| 17 | 4.875 | 1.076 | 37.403 | 5.681 | 1.254 | 36.598 |
| 18 | 4.692 | 1.036 | 38.439 | 5.655 | 1.248 | 37.846 |
| 19 | 4.587 | 1.013 | 39.452 | 5.157 | 1.138 | 38.984 |
| 20 | 4.249 | .938 | 40.390 | 5.121 | 1.130 | 40.115 |
| 21 | 3.997 | .882 | 41.272 | 4.943 | 1.091 | 41.206 |
| 22 | 3.950 | .872 | 42.144 | 4.249 | .938 | 42.144 |

**Table 3 Factor Loadings: PCA Extraction, Quartimax Rotation**

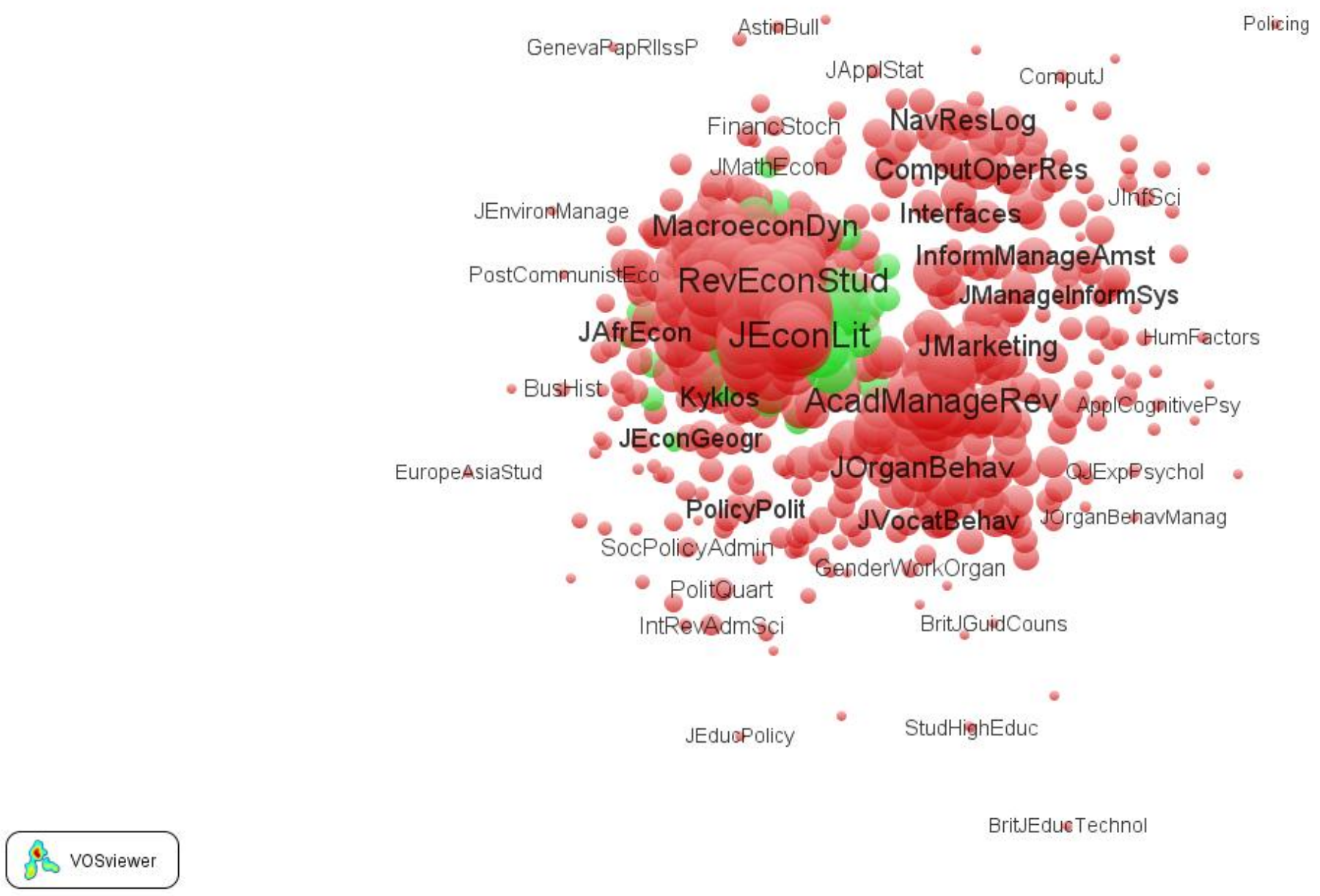

**Figure 1 Grouping of Economics (periphery)**

| **22 Groups** | **No.** | **Mean cites per journal** | **15 Groups** | **No.** | **Mean cites per journal** | **10 Groups** | **No.** | **Mean cites per journal** |
|---|---|---|---|---|---|---|---|---|
| Economics (core) | 56 | 1541 | Economics (core) | 61 | 1282 | Economics (core) | 75 | 1442 |
| Operations research (OR) and operations management | 31 | 1871 | Operations research (OR) and operations management | 33 | 1865 | Operations research (OR) and operations management | 37 | 1653 |
| Management, strategy, SME, sociology | 31 | 2413 | Management, strategy, SME, sociology, *technology* | 42 | 2279 | Management, strategy, SME, sociology, *technology, org psych, IR* | 55 | 1679 |
| HR, org. psychology, org. behaviour | 20 | 1591 | HR, org. psychology, org. behaviour | 24 | 1700 | To management, psychology | | |
| Marketing | 30 | 1523 | Marketing | 31 | 1632 | Marketing | 44 | 1599 |
| Psychology | 19 | 2065 | Psychology | 19 | 2154 | Psychology | 29 | 1627 |
| Finance | 16 | 2105 | Finance | 16 | 2045 | Finance, *Accting* | 26 | 1655 |
| Economics (periphery)* | 20 | 1687 | Economics (periphery)* | 26 | 2032 | Economics (periphery)* | 28 | 1169 |
| Information systems, IT | 23 | 343 | Information systems, IT, *informatics* | 34 | 760 | Information systems, IT, *informatics* | 36 | 609 |
| Regional, environmental studies | 23 | 561 | Regional, environmental studies, *transport* | 31 | 575 | Regional, environmental studies, *transport* | 45 | 631 |
| IR, work, labour | 16 | 403 | IR, work, labour, *labour economics* | 28 | 634 | To HR and Economics | | |
| Energy, environment, agriculture | 16 | 1099 | Energy, environment, agriculture | 16 | 1097 | Energy, environment, agriculture | 23 | 938 |
| Public admin | 18 | 392 | Public admin | 21 | 361 | To psychology and econ peripheral | | |
| Development | 13 | 595 | Development | 12 | 701 | Mainly to economics | | |
| Transport | 9 | 640 | To regional | | | | | |
| Accounting | 10 | 1032 | Accounting | 10 | 1032 | To finance | | |
| Labour econ. | 8 | 859 | To IR | | | | | |
| Technology, Ops Mgt | 17 | 959 | To management, econ, marketing | | | | | |
| Sociology, SME[2] | 15 | 471 | To management, regional, public admin | | | | | |
| Statistics | 12 | 694 | Spread around | | | | | |
| Informatics | 9 | 378 | To IS | | | | | |
| Economic history | 11 | 258 | Spread around | | | | | |
| Total | 423 | | | 374 | | | 348 | |

[1]The economics split is discussed in the text

[2]SME is negatively loaded on to sociology

[3]In the results, the ordering of the groups was different but they have been shown here in alignment to make the comparison easier.

**Table 4 Three possible Sets of Sub-fields in Business and Management**

| **Econ (core)** | **OR.** | **Strategy/Mgt** | **HR** | **Marketing** | **Psychology** | **Finance** | **Econ (periph)** | **IS/IT** | **Regional** | **IR/labour** |
|---|---|---|---|---|---|---|---|---|---|---|
| EUR ECON REV<br>OXFORD B ECON STAT<br>IMF STAFF PAPERS<br>BROOKINGS PAP ECO AC<br>EMPIR ECON<br>J MACROECON<br>REV ECON STAT<br>ECON LETT<br>OPEN ECON REV<br>SCAND J ECON<br>INT J FINANC ECON<br>ECON J<br>ECONOMICA<br>J ECON LIT<br>ECON POLICY<br>J MONETARY ECON<br>APPL ECON LETT | NAV RES LOG<br>EUR J OPER RES<br>OPER RES<br>ANN OPER RES<br>OR SPECTRUM<br>IIE TRANS<br>COMPUT OPER RES<br>OPER RES LETT<br>MANAGE SCI<br>INFORMS J COMPUT<br>J OPER RES SOC<br>INTERFACES<br>J SCHEDULING<br>COMPUT IND ENG<br>PROD PLAN CONTROL<br>INT J PROD ECON<br>M&SOM- | **ADMIN** SCI QUART<br>ACAD MANAGE REV<br>ORGAN SCI<br>ADV STRATEG MANAGE<br>STRATEGIC MANAGE J<br>J MANAGE STUD<br>ACAD MANAGE J<br>ORGAN STUD<br>AM J SOCIOL<br>ANNU REV SOCIOL<br>CALIF MANAGE REV<br>AM SOCIOL REV<br>HUM RELAT<br>INT J MANAG REV<br>ACAD MANAGE PERSPECT | J APPL J. APPL PSYCHOL<br>J ORGAN BEHAV<br>PERS PSYCHOL<br>J OCCUP ORGAN PSYCH<br>J BUS PSYCHOL<br>APPL PSYCHOL-INT REV<br>J OCCUP HEALTH PSYCH<br>EUR J WORK ORGAN PSY<br>HUM PERFORM<br>WORK STRESS<br>J VOCAT BEHAV<br>RES ORGAN BEHAV<br>ORGAN RES METHODS | J MARKETING RES<br>J MARKETING<br>J ACAD MARKET SCI<br>INT J RES MARK<br>EUR J MARKETING<br>MARKET LETT<br>J BUS RES<br>J SERV RES-US<br>J BUS-BUS MARK<br>IND MARKET MANAG<br>J CONSUM RES<br>J BUS IND MARK<br>J RETAILING<br>HARVARD BUS REV<br>MIT SLOAN MANAGE REV<br>J INT MARKETING<br>INT MARKET | PERS SOC PSYCHOL B<br>J PERS SOC PSYCHOL<br>J EXP SOC PSYCHOL<br>PSYCHOL SCI<br>EUR J SOC PSYCHOL<br>PSYCHOL BULL<br>GROUP PROCESS INTERG<br>ANNU REV PSYCHOL<br>PSYCHOL REV<br>BRIT J SOC PSYCHOL<br>J APPL SOC PSYCHOL<br>J BEHAV DECIS MAKING<br>BRIT J PSYCHOL<br>SOCIOL METHODOL | J FINANC QUANT ANAL<br>J FINANC<br>J FINANC ECON<br>REV FINANC STUD<br>J FINANC INTERMED<br>J FINANC MARK<br>FINANC ANAL J<br>J BANK FINANC<br>FINANC MANAGE<br>J CORP FINANC<br>J PORTFOLIO MANAGE<br>EUR FINANC MANAG<br>J LAW ECON<br>J FUTURES MARKETS<br>MATH FINANC | J ECON THEORY<br>GAME ECON BEHAV<br>EXP ECON<br>REV ECON STUD<br>J ECON BEHAV ORGAN<br>ECONOMETRICA<br>THEOR DECIS<br>J PUBLIC ECON<br>J LAW ECON ORGAN<br>ECON THEOR<br>RAND J ECON<br>SOC CHOICE WELFARE<br>J MATH ECON<br>J ECON MANAGE STRAT<br>J RISK | MIS QUART<br>J ASSOC INF SYST<br>INFORM MANAGE-AMSTER<br>INFORM SYST RES<br>EUR J INFORM SYST<br>J MANAGE INFORM SYST<br>J STRATEGIC INF SYST<br>J INF TECHNOL<br>COMMUN ACM<br>INFORM SYST J<br>INFORM SYST MANAGE<br>INT J HUM-COMPUT ST<br>INT J ELECTRON COMM<br>DECIS SUPPORT | REG STUD<br>ECON GEOGR<br>ENVIRON PLANN A<br>URBAN STUD<br>J ECON GEOGR<br>INT J URBAN REGIONAL<br>ANN REGIONAL SCI<br>REG SCI URBAN ECON<br>J URBAN ECON<br>ENVIRON PLANN D<br>EUR PLAN STUD<br>EUR URBAN REG STUD<br>REV INT POLIT ECON<br>J REGIONAL | IND RELAT<br>BRIT J IND RELAT<br>EUR J IND RELAT<br>INT J HUM RESOUR MAN<br>PERS REV<br>IND LABOR RELAT REV<br>INT J MANPOWER<br>ECON IND DEMOCRACY<br>J LABOR RES<br>J WORLD BUS<br>RELAT IND-IND RELAT<br>WORK EMPLOY SOC<br>INT LABOUR REV<br>WORK OCCUPATION<br>GENDER |
| STUD NONLINEAR DYN E<br>OXFORD ECON PAP<br>J MONEY CREDIT BANK<br>MANCH SCH<br>INT ECON REV<br>SCOT J POLIT ECON<br>J INT MONEY FINANC<br>SOUTH ECON J<br>CAN J ECON<br>J INT ECON<br>ECON MODEL<br>J POLIT ECON<br>J ECON PERSPECT<br>APPL ECON<br>ECONOMET J<br>AM ECON REV<br>ECON INQ<br>J ECONOMETRICS<br>ECONOMET | MANUF SERV OP<br>INT J PROD RES<br>OMEGA-INT J MANAGE S<br>TRANSPORT SCI<br>PROD OPER MANAG<br>MATH OPER RES<br>MATH PROGRAM<br>J OPER MANAG<br>EXPERT SYST APPL<br>EXPERT SYST<br>J OPTIMIZ THEORY APP<br>J APPL PROBAB<br>IEEE T SYST MAN CY A<br>INT J COMPUT INTEG M<br>RELIAB ENG SYST SAFE | J INT BUS STUD<br>ORGANIZATION<br>ORGAN DYN<br>J BUS VENTURING<br>LONG RANGE PLANN<br>ENTREP THEORY PRACT<br>BUS ETHICS Q<br>J SMALL BUS MANAGE<br>J BUS ETHICS<br>BRIT J MANAGE<br>J MANAGE INQUIRY<br>INT BUS REV<br>J ORGAN CHANGE MANAG<br>MANAGE LEARN<br>ORGAN ENVIRON<br>ACAD MANAG LEARN EDU | ORGAN BEHAV HUM DEC<br>J MANAGE<br>GROUP ORGAN MANAGE<br>INT J SELECT ASSESS<br>SMALL GR RES<br>LEADERSHIP QUART<br>CAN J ADM SCI<br>BRIT J GUID COUNS<br>GROUP DECIS NEGOT | REV<br>PSYCHOL MARKET<br>MARKET SCI<br>J ADVERTISING<br>J INTERACT MARK<br>J ADVERTISING RES<br>QME-QUANT MARK ECON<br>J PUBLIC POLICY MARK<br>SERV IND J<br>INT J MARKET RES<br>INT J ADVERT<br>SUPPLY CHAIN MANAG<br>TOTAL QUAL MANAG BUS<br>TOURISM MANAGE | PERS INDIV DIFFER<br>GROUP DYN-THEOR RES<br>J EXP PSYCHOL-APPL<br>Q J EXP PSYCHOL<br>APPL COGNITIVE PSYCH | QUANT FINANC | UNCERTAINTY<br>J INST THEOR ECON<br>ECON PHILOS<br>PUBLIC CHOICE<br>J ECON PSYCHOL<br>INT REV LAW ECON | SYST<br>BEHAV INFORM TECHNOL<br>J GLOB INF MANAG<br>INT J INFORM MANAGE<br>INFORM SYST FRONT<br>IEEE T SOFTWARE ENG<br>INTERNET RES<br>INFORM SOFTWARE TECH<br>IND MANAGE DATA SYST<br>ACM T SOFTW ENG METH | SCI<br>J HOUS ECON<br>ENTREP REGION DEV<br>ENVIRON PLANN C<br>J REAL ESTATE FINANC<br>REAL ESTATE ECON<br>J RURAL STUD<br>TIME SOC<br>NEW POLIT ECON<br>CHINA QUART | WORK ORGAN<br>NEW TECH WORK EMPLOY |
| REV<br>CONTEMP ECON POLICY<br>Q J ECON<br>J EUR ECON ASSOC<br>ECON REC<br>MACROECON DYN<br>OXFORD REV ECON POL<br>FISC STUD<br>WORLD ECON<br>J ECON DYN CONTROL<br>REV ECON DYNAM<br>INT TAX PUBLIC FINAN<br>J ECON SURV<br>J POLICY MODEL<br>S AFR J ECON<br>REV WORLD ECON<br>REV INCOME WEALTH<br>J PROD ANAL<br>AM J ECON SOCIOL | | | | | | | | | | |
| KYKLOS<br>DEFENCE<br>PEACE<br>ECON | | | | | | | | | | |
| | | | | | | | | | | |

| Energy | Public Admin | Development | Transport | Accounting | Labour | Technology | Sociology | Statistics | Informatics | Econ. History |
|---|---|---|---|---|---|---|---|---|---|---|
| RESOUR ENERGY ECON<br>ENVIRON RESOUR ECON<br>J ENVIRON ECON MANAG<br>LAND ECON<br>ECOL ECON<br>ENERG J<br>J AGR ECON<br>AUST J AGR RESOUR EC<br>ENERG ECON<br>J REGUL ECON<br>ENERG POLICY<br>EUR REV AGRIC ECON<br>AM J AGR ECON<br>J ENVIRON MANAGE<br>RISK ANAL<br>MAR POLICY | PUBLIC ADMIN<br>PUBLIC MANAG REV<br>POLIT STUD-LONDON<br>GOVERNANCE<br>ADMIN SOC<br>INT REV ADM SCI<br>J PUBL ADM RES THEOR<br>PUBLIC ADMIN REV<br>POLICY POLIT<br>POLIT QUART<br>PUBLIC MONEY MANAGE<br>LOCAL GOV STUD<br>SOC POLICY ADMIN<br>PARLIAMENT AFF<br>J EUR PUBLIC | ECON DEV CULT CHANGE<br>J DEV STUD<br>WORLD DEV<br>WORLD BANK ECON REV<br>J DEV ECON<br>AGR ECON-BLACKWELL<br>WORLD BANK RES OBSER<br>J AFR ECON<br>FOOD POLICY<br>J COMP ECON<br>ECON TRANSIT<br>CHINA ECON REV<br>FEM ECON | TRANSP ORT RES A-POL<br>TRANSPORT REV<br>TRANSPORTATION<br>TRANSPORT POLICY<br>J TRANSP ECON POLICY<br>TRANSPORT RES D-TR E<br>J TRANSP GEOGR<br>TRANSPORT RES B-METH<br>TRANSPORT RES E-LOG | ACCOUNT REV<br>J ACCOUNT RES<br>CONTEMP ACCOUNT RES<br>J ACCOUNT ECON<br>REV ACCOUNT STUD<br>AUDITING-J PRACT TH<br>ACCOUNT ORG SOC<br>EUR ACCOUNT REV<br>J BUS FINAN ACCOUNT<br>ABACUS | J HUM RESOUR<br>LABOUR ECON<br>J LABOR ECON<br>J HEALTH ECON<br>HEALTH ECON<br>REV IND ORGAN<br>J POPUL ECON<br>ECON EDUC REV | RES POLICY<br>R&D MANAGE<br>IND CORP CHANGE<br>INT J TECHNOL MANAGE<br>TECHNOVATION<br>TECHNOL ANAL STRATEG<br>J IND ECON<br>TECHNOL FORECAST SOC<br>IEEE T ENG MANAGE<br>INT J IND ORGAN<br>DECISION SCI<br>J PROD INNOVAT MANAG<br>INT J OPER PROD MAN<br>J EVOL ECON<br>INF ECON POLICY<br>FUTURES<br>TELECOMMUN | SOCIOL OGY<br>SOCIOL REV<br>SOC SCI MED<br>SOCIOL HEALTH ILL<br>BRIT J SOCIOL<br>MILBANK Q<br>ECON SOC<br>SMALL BUS ECON<br>J SOC POLICY<br>CRIT SOC POLICY<br>INT SMALL BUS J<br>THEOR CULT SOC<br>J EUR SOC POLICY<br>HUM ORGAN<br>J LAW SOC | J AM STAT ASSOC<br>J R STAT SOC B<br>ECONOMET THEOR<br>J BUS ECON STAT<br>J R STAT SOC C-APPL<br>J FORECASTING<br>J APPL STAT<br>J R STAT SOC A STAT<br>INT J FORECASTING<br>FINANC STOCH<br>INSUR MATH ECON<br>ASTIN BULL | INFORM PROCESS MANAG<br>J AM SOC INF SCI TEC<br>ANNU REV INFORM SCI<br>J INF SCI<br>INFORM RES<br>INFORM SOC<br>RES EVALUAT<br>INTERACT COMPUT | J ECON HIST<br>ECON HIST REV<br>BUS HIST<br>BUS HIST REV<br>EXPLOR ECON HIST<br>ENTERP SOC<br>HIST POLIT ECON<br>CAMB J ECON<br>EUR J HIST ECON THOU<br>J POST KEYNESIAN EC<br>J ECON ISSUES |
| | POLICY<br>NONPROF<br>VOLUNT<br>SEC Q<br>JCMS-J<br>COMMON<br>MARK S<br>PUBLIC<br>ADMIN<br>DEVELOP | | | | | POLICY | | | | |

**Table 5 Journals in the 22-Group configuration (note that the order of the list does not reflect the quality of the journals)**

| Journals not included in any group (isolates) | Journals that have loadings (>0.1) on 7 or more groups (inter-disciplinary |
|---|---|
| POST-COMMUNIST ECON | REV ECON STAT |
| J MANAGE ENG | J ECON PERSPECT |
| SYST RES BEHAV SCI | EXPERT SYST APPL |
| J CONSTR PSYCHOL | EXPERT SYST |
| J ORGAN BEHAV MANAGE | J SMALL BUS MANAGE |
| ANN TOURISM RES | HARVARD BUS REV |
| J SPORT MANAGE | J LAW ECON |
| NEGOTIATION J | J LAW ECON ORGAN |
| HUM FACTORS | RAND J ECON |
| J RISK INSUR | J ECON MANAGE STRAT |
| INNOV EDUC TEACH INT | J HUM RESOUR |
| TEACH HIGH EDUC | REV IND ORGAN |
| STUD HIGH EDUC | J IND ECON |
| BRIT J EDUC TECHNOL | INT J IND ORGAN |
| EUROPE-ASIA STUD | SMALL BUS ECON |
| J RISK RES | |
| SYST DYNAM REV | |
| SOCIOL TRAV | |
| POLICING | |
| SYST PRACT ACT RES | |
| ERGONOMICS | |
| BRIT EDUC RES J | |
| J ADV NURS | |
| J EDUC POLICY | |
| PHYSICA A | |
| GENEVA PAP R I-ISS P | |
| J HIGH EDUC | |
| IEEE T INF TECHNOL B | |

**Table 6 Journals that are isolated from others and journals that are inter-disciplinary**